\documentclass[12pt]{iopart}
\usepackage{cite}
\usepackage{graphicx}
\usepackage{iopams}
\usepackage{bm}

\usepackage[usenames,dvipsnames,svgnames,table]{xcolor}
\usepackage[unicode=true,
            pdfusetitle,
            bookmarks=true,
            bookmarksnumbered=false,
            bookmarksopen=false,
            breaklinks=true,
            pdfborder={0 0 0},
            backref=false,
            colorlinks=true]{hyperref}
\hypersetup{linkcolor=NavyBlue,urlcolor=NavyBlue,citecolor=NavyBlue}

\providecommand{\openone}{\leavevmode\hbox{\small1\kern-4.3pt\normalsize1}}
\renewcommand\Re{\mathrm{Re}}

\begin{document}

\title{Quantum multiparameter metrology with generalized entangled coherent state}

\author{Jing Liu}
\address{Department of Physics, Zhejiang
University, Hangzhou 310027, China}
\address{Department of Mechanical and Automation Engineering, The 
Chinese University of Hong Kong, Shatin, Hong Kong}

\author{Xiao-Ming Lu}
\address{Department of Electrical and Computer Engineering, National University
of Singapore, 4 Engineering Drive 3, Singapore 117583}

\author{Zhe Sun}
\address{Department of Physics, Hangzhou Normal University, Hangzhou 310036, China}

\author{Xiaoguang Wang}
\address{State Key Laboratory of Modern Optical Instrumentation, 
Department of Physics, Zhejiang
University, Hangzhou 310027, China}
\address{Synergetic Innovation Center of Quantum Information  
and Quantum Physics, University of Science and Technology of China, 
Hefei, Anhui 230026, China}
\ead{xgwang@zimp.zju.edu.cn}

\begin{abstract}
We propose a generalized form of entangled coherent states (ECS) and apply them in a multi-arm optical interferometer to estimate 
multiple phase shifts. We obtain the quantum Cram\'er-Rao bounds for both 
the linear and nonlinear parameterization protocols.  Through the analysis, we find that, 
utilizing the simultaneous estimation, this generalized form of ECS gives a better precision than the generalized  NOON states 
{[}Phys. Rev. Lett. \textbf{111}, 070403 (2013){]}. Moreover, comparing with the independent estimation, 
both the linear and nonlinear protocols have the same advantage in the relation to the number of the parameters. 
\end{abstract}

\pacs{03.67.-a, 06.90.+v, 42.50.Dv, 42.50.St}

\maketitle

\section{Introduction}

Since the pioneer work of Caves in 1981~\cite{Caves}, quantum metrology
has made a great progress as a successful application of quantum mechanics to enhance the measurement precision~\cite{intro0,intro00,intro1,intro10,intro11,intro2,
intro20,intro21,intro22,Lu,Lu1,intro3,intro31,intro32}.
However, unlike the single-parameter quantum metrology, multiparameter quantum metrology for a long time was not adequately studied. One major reason is that quantum multiparameter
Cram\'{e}r-Rao bound in general cannot be saturated. 
A decade ago, the condition of this bound to be tight for pure states has been given~\cite{Matsumoto,Fujiwara}. Since then, several protocols
on multiparameter estimation were proposed in different scenarios~\cite{Walmsley,Yue,Pinel,Yao,Yao1,lang}.
One of these works was given by Humphreys \emph{et al.} in Ref.~\cite{Walmsley},
where the phase imaging problem was mapped into a multiparameter metrology
process and a generalized form of NOON states was used as the input
resource. 
They found that the simultaneous estimation with the generalized NOON states is better than the independent estimation
with the NOON state.

On the other hand, the NOON state is not the only state that is available
to reach the Heisenberg limit. Another useful state is the so-called
entangled coherent state (ECS), which has been widely applied and
studied in quantum metrology recently~\cite{Sanders2012,GRJin,Liupra,Joo,ECS,Gerry}.
In a Mach-Zehnder interferometer, ECS has been proved to be more powerful
than NOON state in giving a Heisenberg scaling precision~\cite{Joo}.
Even in a lossy interferometer, ECS can still beat the shot-noise
limit for a not very large loss rate~\cite{GRJin,Liupra}. Thus,
it is reasonable for one to wonder that if a generalized form of ECS
could give a better theoretical precision than the counterpart of
NOON state. This is the major motivation of this work.

In this paper, we apply a generalized form of ECS in linear and nonlinear
optical interferometers. By calculating the quantum Fisher information
matrix (QFIM), we give the analytical expression of the quantum Cram\'{e}r-Rao bound (QCRB)
for both linear and nonlinear protocols. In the linear protocol, the
bound can reach the Heisenberg scaling for most values of the total
photon number. Meanwhile, with respect to the parameter number $d$,
in both protocols, for most values of photon number, the bounds provide 
better precisions than that given by independent protocol
with ECS or NOON state, which is the same as the generalized NOON
state discussed in Ref.~\cite{Walmsley}.


The paper is organized as follows. In Sec.~\ref{sec:Cramer-Rao-bound},
we briefly review the quantum Cram\'{e}r-Rao bound as well as the quantum Fisher information matrix for multiparameter estimations. 
In Sec.~\ref{sec:ECS}, we introduce a generalized form of entangled coherent state for multiple modes  and apply it 
in linear and nonlinear optical interferometers. 
Furthermore, the comparison between this state and the generalized NOON state is
discussed. In Sec.~\ref{sec:Measurement}, we discuss the  optimal 
measurement problem to achieve the bound. 
In Sec.~\ref{random}, we extend our discussion to random variables and compare the generalized ECS 
and NOON state with quantum Ziv-Zakai bound. Section~\ref{sec:Conclusion}
is the conclusion.

\section{Quantum Cram\'{e}r-Rao bound} \label{sec:Cramer-Rao-bound}

In a multiparameter quantum metrology process, the quantum state $\rho$ depends on a set of deterministic parameters $\bm\theta=\{\theta_j\}$. 
The value of $\bm\theta$ is estimated by processing the observation data obtained by measuring the quantum system.
A generalized quantum measurement is characterized by a positive-operator-valued measure $\{E_x\}$ with $x$ denoting outcomes.
According to quantum mechanics, the probability of obtaining the outcome $x$ is $p(x)=\mathrm{Tr}(E_x\rho)$.
Denote the estimator for $\theta_j$ by $\hat\theta_j$, which is a map from the measurement outcome $x$ to the estimates.
The accuracy of the multiparameter estimation can be measured by the estimation-error covariance matrix:
$C_{jk}:=\int\!dx\,[\hat\theta_j(x)-\theta_j][\hat\theta_k(x)-\theta_k]\mathrm{Tr}(E_x\rho)$.
For (locally) unbiased estimators $\hat\theta_j$, the QCRB on the estimation error reads~\cite{Helstrom,Holevo}
\begin{equation}
    C\geq(\nu \mathcal{F})^{-1},\label{eq:C_R}
\end{equation}
where $\nu$ is the number of the repetition of the experiments, and $\mathcal{F}$ is the quantum Fisher information matrix (QFIM).
Let $L_j$ be the symmetric logarithmic derivative (SLD) for $\theta_j$, which is a Hermitian operator satisfying $\partial\rho/\partial\theta_j=\left(\rho L_j+L_j\rho\right)/2$.
Then, the QFIM is defined by
\begin{equation}
\mathcal{F}_{jk}=\frac12\mathrm{Tr}\left[(L_j L_k + L_k L_j)\rho \right].
\end{equation}
Recently, it has been found that similarly with the quantum Fisher
information~\cite{Liu2013}, the QFIM can also be expressed in the
support of the density matrix~\cite{liu2014}. 
Denote $\partial_j:=\partial/\partial \theta_j$ for simplicity henceforth.
For a pure state $\rho=|\psi\rangle\langle\psi|$, the elements of QFIM can be expressed as~\cite{Helstrom,Holevo}
\begin{equation}
\mathcal{F}_{jk}=4\mathrm{Re}\left(
    \langle\partial_j\psi|\partial_k\psi\rangle-
    \langle\partial_j\psi|\psi\rangle\langle\psi|\partial_k\psi\rangle
\right).
\end{equation}

In this work, we use the the total variance $|\delta\hat{\bm\theta}|^{2}:=\mathrm{Tr}\,C$ as a figure of merit for the multiparameter estimation. 
Taking the trace on both sides of inequality (\ref{eq:C_R}) leads to
\begin{equation}
|\delta\bm{\hat{\theta}}|^{2}\geq\mathrm{Tr}(\mathcal{F}^{-1}),\label{eq:tot_var}
\end{equation}
where we have set $\nu=1$ as we are only interested in the quantum enhancements. 
For a two-parameter case, Eq.~(\ref{eq:tot_var}) is reduced into $|\delta\bm{\hat{\theta}}|^{2}\geq1/F_{e}$,
where $F_{e}=\mathrm{det}\mathcal{F}/\mathrm{Tr}\mathcal{F}$ with $\mathrm{det}(\cdot)$ denoting the determinant can
be treated as an effective quantum Fisher information. 

In order to draw conclusion on the best possible estimation error from the QCRB, it is important to know whether the lower bound is achievable. 
The QCRB for multiparameter estimation is in general not achievable. 
However, for  pure states, the QCRB can be saturated if $\mathrm{Im}\langle\psi|L_j L_k|\psi\rangle=0$ are satisfied for all $j$, $k$, and $\bm\theta$~\cite{Matsumoto,Fujiwara,Walmsley}. 
Note that $L_j=2\partial_j(|\psi\rangle\langle\psi|)$ is an SLD operator for $\theta_j$.
It can be shown that $\mathrm{Im}\langle\psi|L_j L_k|\psi\rangle=0$ is equivalent to
$\langle\partial_j \psi|\partial_k \psi\rangle\in\mathbb{R}$.
For a unitary parametrization process,  i.e.,  
$|\psi\rangle=U_{\bm\theta}|\psi_{\mathrm{in}}\rangle$,  this condition 
can be rewritten as~\cite{single2}
\begin{equation}
\langle \psi_{\mathrm{in}}|[\mathcal{H}_j,\mathcal{H}_k]
    |\psi_{\mathrm{in}}\rangle=0,  \quad\forall\, j,k,
\end{equation}
where $\mathcal{H}_j:=i(\partial_j U_{\bm \theta}^{\dagger})U_{\bm\theta}$ is the characteristic operator for the parameterization of the $l$th parameter.


\begin{figure}[t]
\center\includegraphics[width=9cm]{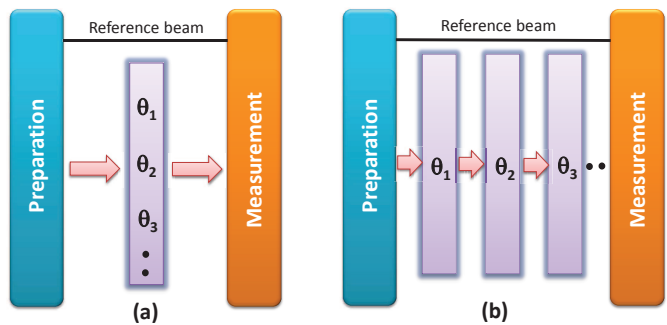}\protect
\caption{\label{fig:parallel}
Two methods to sense multiple parameters: (a) the simultaneous protocols and (b) the sequential protocols.}
\end{figure}

The single-parameter unitary parametrization processes have been detailedly 
discussed in recent works~\cite{single2,single1,single3}.
For a multiparameter  unitary parametrization process, there are two basic methods to sense the multiple parameters: the simultaneous protocols and the sequential protocols, as shown in Fig.~\ref{fig:parallel}. 
Let $\{H_j\}$ be a set of Hermitian operators.
Then, the simultaneous sensing is described by 
\begin{equation}
    U_{\mathrm{I}}=\exp\left(\sum_{j=1}^{d}i H_j\theta_j\right)
\end{equation}
with $d$ being the number of the parameters, while the sequential sensing is described by
\begin{equation}
U_{\mathrm{II}}=\prod_{j=1}^{d}\exp(iH_{d-j+1}\theta_{d-j+1}).
\end{equation}

In this  paper, we focus on the phase estimations in the optical multi-arm interferometer,  
in which $H_j$ is a local operator on the $j$th mode. 
Thus, all the generating operators $H_j$ are commutative, and these two methods 
are equivalent. 
Moreover, in such a case, it is easy to show that $\mathcal{H}_j=H_j$, and thus 
$\langle\psi_{\mathrm{in}}|[\mathcal{H}_j, \mathcal{H}_k]|\psi_{\mathrm{in}}\rangle
=\langle\psi_{\mathrm{in}}|[H_k,H_j]|\psi_{\mathrm{in}}\rangle=0$.
This implies that the Cram\'{e}r-Rao bound is theoretically achievable.  
The element of QFIM can be written as
\begin{equation}
    \mathcal{F}_{jk} = 4\Big(
    \langle\psi_{\mathrm{in}}|H_j H_k|\psi_{\mathrm{in}}\rangle
    -\langle\psi_{\mathrm{in}}|H_j|\psi_{\mathrm{in}}\rangle\langle\psi_{\mathrm{in}}|H_k|\psi_{\mathrm{in}}\rangle
    \Big).
\label{eq:f}
\end{equation}

\section{Generalized entangled coherent state for multiparameter estimation}\label{sec:ECS}

\subsection{Generalized entangled coherent state}

Entangled coherent state (ECS) has been applied in several single-parameter
quantum metrology protocols and proved to be powerful in beating
the shot-noise limit~\cite{GRJin,Liupra,Joo}. 
Let $|\alpha\rangle$ be a coherent state and $|0\rangle$ the vacuum state.
The ECS is given by~\cite{ECS,Gerry}
\begin{equation}
|\mathrm{ECS}\rangle=\mathcal{N}\left(|\alpha 0\rangle+|0\alpha\rangle\right),
\end{equation}
where $\mathcal{N}=[2(1+e^{-|\alpha|^{2}})]^{-1/2}$ is the normalizing factor. 
Taking the ECS as the input, the QCRB for a parameter sensed by the Hamiltonian $H=a^\dagger a$ with $a$ denoting the annihilation operator of the first mode is given by~\cite{Joo}
\begin{equation}
    |\delta\bm{\hat{\theta}}|_{\mathrm{ECS}}^{2}=\frac{1}{4\mathcal{N}^{2}|\alpha|^{2}
\left[1+|\alpha|^{2}\left(1-\mathcal{N}^{2}\right)\right]}.
\end{equation}
Therefore, for the independent estimations for $d$ parameters, the QCRB on the total variance is 
$|\delta\bm{\hat{\theta}}|_{\mathrm{ind}}^{2}=d|\delta\bm{\hat{\theta}}|_{\mathrm{ECS}}^{2}$.
In terms of the mean number of total photons involved, i.e., $N_{\mathrm{tot}}=2d\mathcal{N}^{2}|\alpha|^{2}$, we obtain
\begin{equation}
|\delta\bm{\hat{\theta}}|_{\mathrm{ind}}^{2}=\frac{d^{3}}{N_{\mathrm{tot}}
\left[2d+N_{\mathrm{tot}}\left(\mathcal{N}^{-2}-1\right)\right]}.
\end{equation}
When $|\alpha|$ is large, $\mathcal{N}^{-2}\rightarrow2$, the total
variance
\begin{equation}
|\delta\bm{\hat{\theta}}|_{\mathrm{ind}}^{2}\rightarrow\frac{d^{3}}
{N_{\mathrm{tot}}\left(N_{\mathrm{tot}}+2d\right)}.
\end{equation}
This bound is lower than $d^{3}/N_{\mathrm{tot}}^{2}$, which is given by NOON state in the independent estimation~\cite{Walmsley}.

Multiple parameters can also be sensed and estimated by entangled input states, e.g., using a multi-arm interferometer with generalized NOON states as input~\cite{Walmsley}.
Here, we use a generalized form of the ECS instead, as the ECS is more powerful than the NOON states in the two-arm interferometer for a fixed mean number of the total photons. 
For a multiparameter estimation scenario as shown in Fig.~\ref{fig:parallel}, we set the reference beam as mode zero and the parametrized beams as mode $1$ to mode $d$.
Taking into consideration the symmetry among $d$ modes, we generalize the ECS to the multi-mode case as 
\begin{equation}
    |\psi\rangle=b\sum_{j=1}^d|\alpha\rangle_j + c|\alpha\rangle_0,\label{eq:in}
\end{equation}
where $|\alpha\rangle_j=\exp(\alpha a^{\dagger}_j-\alpha^{*}a_j)|0\rangle$ with $|0\rangle$ being the multi-mode vacuum is a state with a coherent state in $j$th mode and vacuums in others. 
The coefficients $b$ and $c$ are complex numbers; due to the normalization of $|\psi\rangle$, they satisfy
\begin{equation}\label{eq:normalization}
    |c|^2 + (bc^*+cb^*)v + |b|^2 u = 1
\end{equation}
with 
\begin{equation}\label{eq:uv}
    u:=d+d(d-1)e^{-|\alpha|^2}  \mbox{ and } v:=de^{-|\alpha|^2}. 
\end{equation}
In this paper, we will use this generalized form of ECS as the input state to sense the parameters.


\subsection{Local parameterization}
Let us consider that the parameters are sensed via $U_{\bm\theta}=\exp(i\sum_{j=1}^d H_j \theta_j)$ with $H_j=(a_j^{\dagger}a_j)^m$, where $m$ is a positive integer.
Taking the generalized ECS Eq.~(\ref{eq:in}) as the input state, it follows that 
\begin{equation}
    \langle\psi|H_j|\psi\rangle = |b|^2f(m,\alpha)
    \quad\mbox{and}\quad
    \langle\psi|H_j H_k|\psi\rangle = |b|^2 f(2m,\alpha)\delta_{jk}
\end{equation}
with $f(m,\alpha) := \langle \alpha|(a^\dagger a)^m|\alpha\rangle$.
From Eq.~(\ref{eq:f}), the elements of the QFIM are given by
\begin{equation}
    \mathcal{F}_{jk} = 4 [\delta_{jk} |b|^2 f(2m,\alpha) - |b|^4 f(m,\alpha)^2].
\end{equation}
Consequently, the QFIM can be expressed as
\begin{equation}
    \mathcal{F}=4 |b|^2 f(2m,\alpha)
    \left(\openone - \frac{|b|^2 f(m,\alpha)^2}{f(2m,\alpha)}\mathcal{I}\right),
\end{equation}
where $\openone$ is the identity matrix, and $\mathcal{I}$ is the
matrix with elements $\mathcal{I}_{jk}=1$ for
all $j$ and $k$. 
Noting that $\mathcal{I}^2=d\mathcal{I}$, it can be shown that 
\begin{equation}
    \left[\gamma(\openone+\omega\mathcal{I})\right]^{-1} 
    = \frac{1}{\gamma}\left(\openone-\frac{\omega}{1+\omega d}\mathcal{I}\right)
\end{equation}
with $\gamma$ and $\omega$ being real numbers.
Thus, we obtain the analytical result for the inverse of the QFIM as
\begin{equation}
    \mathcal{F}^{-1} = \frac{1}{4|b|^2 f(2m,\alpha)}
    \left(\openone+
    \frac{|b|^2f(m,\alpha)^2}{f(2m,\alpha)-|b|^2f(m,\alpha)^2d}
    \mathcal{I}\right).
\end{equation}
Tracing both sides of the above equality, we obtain the lower bound on the total variance:
\begin{equation}\label{eq:general_bound}
    |\delta\hat{\bm\theta}|^2 
    \geq
    \mathrm{Tr}(\mathcal{F}^{-1})=
    \frac{d}{4f(2m,\alpha)}\left(\frac{1}{|b|^2}+\frac{1}{g-|b|^2d}\right)
\end{equation}
with $g:=f(2m,\alpha)/f(m,\alpha)^2$ being a nonnegative number.

Since we are interested in the minimal estimation error, we minimize the QCRB on the total variance over $b$, which is equivalent to minimize the quantity $w:=1/|b|^2+1/(g-|b|^2d)$ in Eq.~(\ref{eq:general_bound}) over $b$.
By noting that $|b|^2$ takes value in a continuous range and $\mathrm{Tr}(\mathcal{F}^{-1})$ is nonnegative, it follows from Eq.~(\ref{eq:general_bound}) that we only need to investigate those $|b|^2<g/d$, as $\mathrm{Tr}(\mathcal{F}^{-1})$ will be negative when $|b|^2$ passes through $g/d$. 
In this domain, the minimum of $w$ is at the place where the derivative of $w$ with respect to $|b|^2$ is zero, that is
\begin{equation}
    |b|=b_\star:= \sqrt{g/(\sqrt{d} + d)}.   
\end{equation}
Note that the rigor domain of $|b|^2$ is determined by the normalization condition Eq.~(\ref{eq:normalization}).
Due to an irrelevant global phase in the states in Eq.~(\ref{eq:in}), we can always assume that $c$ is real.
Equation (\ref{eq:normalization}) then becomes
\begin{equation}
    c^2 + 2(\Re\,b)vc + |b|^2 u  - 1 = 0,
\end{equation}
which has a solution for $c$ only if  $(\Re\,b)^2 v^2 - |b|^2 u + 1 \geq 0$.
Since $|b|\geq|\Re\,b|$, we obtain
\begin{equation}
    |b|^2 \leq\frac1{u-v^2} = \frac1{d+d(d-1)e^{-|\alpha|^2}-d^2e^{-2|\alpha|^2}}\equiv\Gamma,
\end{equation}
which implies that the domain of $|b|^2$ is $[0,\Gamma]$. 
Therefore, we obtain
\begin{equation}\label{eq:optimal}
    \min_{|b|\in[0,\sqrt\Gamma]}\mathrm{Tr}(\mathcal{F}^{-1})=
\cases{
    \frac{d(1+\sqrt d)^2}{4}\frac{f(m,\alpha)^2}{f(2m,\alpha)^2}& for  $b_\star\leq\sqrt\Gamma$ \\
    \frac{d}{4f(2m,\alpha)}\left(\frac{1}{\Gamma}+\frac{1}{g-\Gamma d}\right) & for $b_\star>\sqrt\Gamma$.
}
\end{equation}

\begin{figure}
\center\includegraphics[width=10cm,height=5cm]{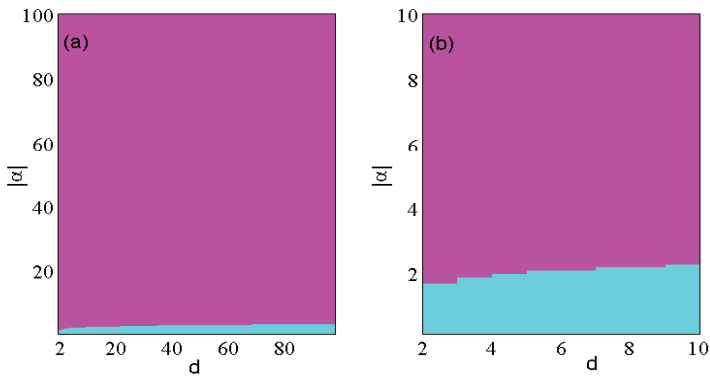}
\caption{\label{fig:b_alpha}
Region partition according to whether $b_\star$ is in the domain of $b$ for the linear parameterization with $H_j=a_j^\dagger a_j$.
The blue region represent where $b_\star$ is outside the domain of $b$.
}
\end{figure}

We first consider the linear parametrization protocol, for which $m=1$. 
Note that $f(1,\alpha)=|\alpha|^2$, $f(2,\alpha)=|\alpha|^2(1+|\alpha|^2)$, and therefore $g=1+|\alpha|^{-2}$.
From Eq.~(\ref{eq:optimal}), we obtain the lower bound on the total variance:
\begin{equation}\label{eq:general_bound1}
    |\delta\hat{\bm\theta}|^2 
    \geq
    \mathrm{Tr}(\mathcal{F}_{\mathrm{L}}^{-1})=
    \frac{d}{4|\alpha|^2(1+|\alpha|^2)}
    \left(
        \frac1{|b|^2} + \frac{1}{1+|\alpha|^{-2} -d |b|^2}
    \right).
\end{equation}
After minimizing over $b$, we obtain 
\begin{equation}\label{eq:CR_linear}
|\delta\bm{\hat{\theta}}|^2\geq|\delta\bm{\hat{\theta}}|_{\mathrm{L}}^2
=\frac{d(\sqrt{d}+1)^2}{4\left(1+|\alpha|^2\right)^2},
\end{equation}
provided that $b_\star\leq\sqrt\Gamma$.
Figure~\ref{fig:b_alpha} shows the parameter regime where $|b_\star|\leq\sqrt\Gamma$.
The purple areas in both panels represent the regime where
$b_\star$ is inside the domain of $b$.
From Fig.~\ref{fig:b_alpha}(a),
it can be found that $b$ can be $b_\star$ for a large $|\alpha|$. 
This is reasonable because when $|\alpha|$ is infinite, $b_\star=1/\sqrt{d+\sqrt{d}}$ and $\Gamma=1/d$,
$b_\star$ is always less than $\sqrt\Gamma$. 
For a large $d$, $b_\star$ may be beyond the domain of $b$ when $|\alpha|$ is very small.
A more experimentally realizable regime is
that both of $d$ and $|\alpha|$ are not very large and comparable
to each other, as shown in Fig.~\ref{fig:b_alpha}(b). In this regime,
$b_\star$ is inside the domain of $b$ for most areas. Especially, when
$|\alpha|$ is larger than around $2$, $b_\star$ is always
reachable, indicating that the corresponding bound $|\delta\bm{\hat{\theta}}|_{\mathrm{L}}^{2}$
can always be reached.

\begin{figure}
\center\includegraphics[width=10cm,height=5cm]{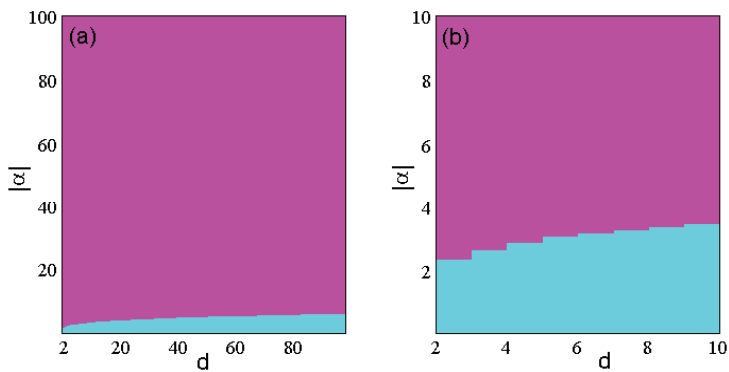}
\protect\caption{\label{fig:b_n_alpha}
Region partition according to whether $b_\star$ is in the domain of $b$ for the nonlinear parameterization with $H_j=(a_j^\dagger a_j)^2$.
The blue region represent where $b_\star$ is outside the domain of $b$.
}
\end{figure}

We also consider a nonlinear parametrization protocol for which $m=2$. 
It can be shown that $f(2,\alpha)=|\alpha|^2(1+|\alpha|^2)$ and $f(4,\alpha)=|\alpha|^{8}+6|\alpha|^{6}+7|\alpha|^{4}+|\alpha|^{2}$.
After minimizing over $b$, we obtain 
\begin{equation}
|\delta\bm{\hat{\theta}}|^2\geq|\delta\bm{\hat{\theta}}|_{\mathrm{NL}}^2
=\frac{d(\sqrt{d}+1)^2}{4}\left(\frac{1+|\alpha|^2}{|\alpha|^{6}+6|\alpha|^{4}+7|\alpha|^{2}+1}\right)^2,
\end{equation}
provided that $b_\star\leq\sqrt\Gamma$.
In Fig.~\ref{fig:b_n_alpha} we can see that, similarly to Fig.~\ref{fig:b_alpha}, $b_\star\leq\sqrt\Gamma$ is satisfied in most areas.

\subsection{Analysis}
Here, we give an analysis on the QCRB given by the generalized ECS, and compare them with the ones given by the generalized NOON states, which was applied in multiparameter metrology in Ref.~\cite{Walmsley}.
The generalized  NOON states proposed in Ref.~\cite{Walmsley} reads $|\psi_{\mathrm{s}}\rangle=b\sum_{j=1}^d|N\rangle_j+c|N_0\rangle$, where $|N\rangle_j=|\cdots00N00\cdots\rangle$ is the state with $N$ photons in $j$th mode and vacuum in others. 
For the linear parametrization protocol $H=\sum_{j=1}^d a_j^{\dagger} a_j$, the minimal QCRB on the total variance over all generalized NOON states for a given $N$ is~\cite{Walmsley}
\begin{equation}
|\delta\bm{\hat{\theta}}|_{\mathrm{sL}}^{2}=\frac{d(\sqrt{d}+1)^{2}}{4N^{2}},
\end{equation}
where the optimal value of $b$ is $b=1/\sqrt{d+\sqrt{d}}$. 
For the nonlinear parametrization protocol $H=\sum_{j=1}^d (a_j^{\dagger}a_j)^2$, through
some straightforward calculations, we obtain the minimal QCRB 
\begin{equation}
|\delta\bm{\hat{\theta}}|_{\mathrm{sNL}}^{2}=\frac{d(\sqrt{d}+1)^{2}}{4N^{4}},
\end{equation}
which is also attained at $b=1/\sqrt{d+\sqrt{d}}$.

From the expressions of $|\delta\bm{\hat{\theta}}|_{\mathrm{L}}^{2}$,
$|\delta\bm{\hat{\theta}}|_{\mathrm{NL}}^{2}$, $|\delta\bm{\hat{\theta}}|_{\mathrm{sL}}^{2}$
and $|\delta\bm{\hat{\theta}}|_{\mathrm{sNL}}^{2}$, we find 
that all these bounds share the same scaling relation with respect to the parameter number $d$; they are all proportional
to $d(\sqrt{d}+1)^{2}$. 
Furthermore, both $|\delta\bm{\hat{\theta}}|_{\mathrm{L}}^{2}$ and $|\delta\bm{\hat{\theta}}|_{\mathrm{NL}}^{2}$
provide a $\mathcal{O}(d)$ advantage compared to the independent
estimation with ECS or NOON state, 
which is the same as $|\delta\bm{\hat{\theta}}|_{\mathrm{sL}}^{2}$~\cite{Walmsley}.

These protocols show different relations to the average total photon number $N_{\mathrm{tot}}$. 
Obviously, the average total
photon number of $|\psi_{\mathrm{s}}\rangle$ is $N_{\mathrm{s,tot}}=N$. 
Meanwhile, the average total photon number of $|\psi_{\alpha}\rangle$ is
$N_{\alpha,\mathrm{tot}}=|\alpha|^2\left(d|b|^2+|c|^2\right)$,
which is dependent on the values of $b$ and $c$. 
When $\alpha$ is sufficiently large such that $d\exp(-|\alpha|^2)\ll1$, we have $N_{\alpha,\mathrm{tot}}\simeq|\alpha|^2$ as a result of  $d|b|^2+|c|^2\simeq1$ implied by the normalization condition Eq.~(\ref{eq:normalization}).
As a matter of fact, when $|\alpha|=4$,
$\exp(-|\alpha|^{2})\simeq10^{-7}$. For a not very large $d$, choosing
$d=5$ for example, at the optimal value $b_\star$ of $b$, the difference between $N_{\mathrm{\alpha,tot}}$ and $|\alpha|^{2}$ is around $10^{-6}$.
Thus, for most values of $|\alpha|$ and $d$, the average photon
number can be approximated as $|\alpha|^{2}$. With this approximation,
$|\delta\bm{\hat{\theta}}|_{\mathrm{L}}^{2}\varpropto N_{\mathrm{\alpha,tot}}^{-2}$
and $|\delta\bm{\hat{\theta}}|_{\mathrm{NL}}^{2}\varpropto N_{\mathrm{\alpha,tot}}^{-4}$.

\begin{figure}
\center\includegraphics[width=8cm]{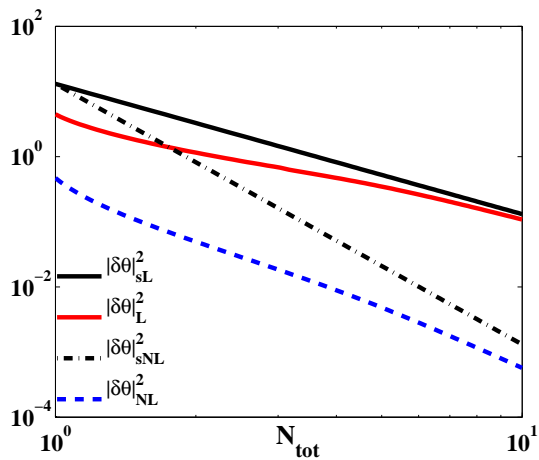}
\protect\caption{\label{fig:N_tot}
Variation of $|\delta\bm{\theta}|_{\mathrm{sL}}^{2}$,
$|\delta\bm{\theta}|_{\mathrm{L}}^{2}$, $|\delta\bm{\theta}|_{\mathrm{sNL}}^{2}$
and $|\delta\bm{\theta}|_{\mathrm{NL}}^{2}$ as functions of the
average total photon number $N_{\mathrm{tot}}$. 
The black and red
solid lines represent $|\delta\bm{\theta}|_{\mathrm{sL}}^{2}$ and
$|\delta\bm{\theta}|_{\mathrm{L}}^{2}$, the QCRB
with the generalized NOON states and the generalized ECS for
linear protocol, respectively. 
The black dash-dot and blue dash lines
represent $|\delta\bm{\theta}|_{\mathrm{sNL}}^{2}$ and $|\delta\bm{\theta}|_{\mathrm{NL}}^{2}$,
the counterpart with with the generalized NOON states and the generalized ECS 
for nonlinear protocol, respectively. 
The total parameter number is set to $d=5$ here.}
\end{figure}

In Fig.~\ref{fig:N_tot}, we plot these four QCRBs, 
$|\delta\bm{\hat{\theta}}|_{\mathrm{L}}^{2}$,
$|\delta\bm{\hat{\theta}}|_{\mathrm{NL}}^{2}$, 
$|\delta\bm{\hat{\theta}}|_{\mathrm{sL}}^{2}$,
and $|\delta\bm{\hat{\theta}}|_{\mathrm{sNL}}^{2}$ as functions of the average total photon number. 
Comparing $|\delta\bm{\hat{\theta}}|_{\mathrm{sL}}^{2}$
and $|\delta\bm{\hat{\theta}}|_{\mathrm{L}}^{2}$ in linear protocols,
in the regime of small average total photon numbers, even $b_\star$ may be greater than $\sqrt\Gamma$, the generalized ECS still gives a lower QCRB than the generalized NOON states. 
However, this advantage reduces when $N_{\mathrm{tot}}$ increases.  
For a very large $N_{\mathrm{tot}}$, the generalize ECS and the generalized NOON states are basically equivalent to each other on the estimation precision. 
Besides, the nonlinear parametrization process is always better than the
linear one for the same input state in this case, as expected. 
What is more interesting here is that, for a small $N_{\mathrm{tot}}$, the linear
protocol with generalized ECS can give a lower bound than
the nonlinear counterpart with generalized NOON states. 
This gives an alternative strategy for small photon number scenario when the nonlinear parametrization is very challenging to perform.

\section{Measurement} \label{sec:Measurement}

For an entire metrology process, the measurement has to be considered
as the QCRB cannot be always saturated for any
measurement. As a matter of fact, different measurement strategies
would give different classical Cram\'{e}-Rao bound and further give different
metrology scalings.

For the estimation of a single parameter, the projective measurement with respect to the eigenstates of the SLD operator can be 
used as the optimal measurement if they are (locally) independent of the parameter under estimation~\cite{Caves94,Zhong14,Paris}.
For the cases where the eigenstates of the SLD operator depend on the true value of the parameter, one has to resort to the adaptive measurement and estimation scheme to asymptotically attain the QCRB~\cite{Nagaoka1988,Fujiwara2006}.
For multiparameter estimations, due to the non-commutativity of the SLDs, the QCRB in general cannot be attained.  
However, for multiparameter estimation with pure states, the QCRB can be attained if $\mathrm{Im}\langle\psi|L_j L_k|\psi\rangle=0$ for all $j$, $k$, and $\bm\theta$~\cite{Matsumoto,Fujiwara}, which is satisfied for the case consider in this work.
In principle, there exists an optimal measurements asymptotically attain the QCRB, although this optimal measurement may be hard to  implemented experimentally.
General methods to construct such an optimal measurement can be found in Ref.~\cite{Matsumoto, Walmsley}.

\section{Deterministic parameter versus random parameter} \label{random}

During the entire calculation of the paper, we treat the phase shifts as unknown but deterministic 
signals~\cite{Levy,Poor,Lehmann,Rivas},  which means that the true values of phase shifts are always the same during the repetitions of the experiment.
In other words,  we have independent and identically distributed samples to perform measurements, with the collection of the measurement outcomes the parameters are estimated. 
For example,  during the detection of the gravity, the gravity is commonly treated as a deterministic parameter. Meanwhile, 
optical interferometry  is a major approach for this detection,  for example the LIGO (Laser Interferometer 
Gravitational-Wave Observatory) program.  Thus, it is reasonable to treat the phase shifts as deterministic 
signals.
 
However, in some different scenarios, for instance, during the measure of the gravitational acceleration in 
a specific location on earth,  its value may be slightly affected by the flow of some underground magma or 
geology movement. Thus,  it is also reasonable to treat the signal as a a random parameter in these scenarios.
Recently,  Tsang proposed a quantum version of Ziv-Zakai bounds for estimating a random parameter~\cite{Tsang}. 
Using this bound,  Giovannetti and Maccone found that for high prior information regime, the accuracy given by 
sub-Heisenberg strategies is no better than that obtained by guessing according to the prior distribution~\cite{Giovannetti}. 
Thus, the precision for a random variable and  a deterministic parameter may have great differences.
For the generalized NOON state, the quantum Ziv-Zakai bound has been given by Zhang and Fan in 
Ref.~\cite{zhang} as
\begin{equation}
|\delta\bm{\hat{\theta}}|^2\geq\mathrm{max}\left\{\frac{d(d+\sqrt{d})^2 }{80\lambda^2 N^2}, 
\frac{(\pi^2/16-0.5)d(d+\sqrt{d})^2}{(d+\sqrt{d}-1)N^2}\right\} \label{eq:ZZ_NOON}
\end{equation}
with $\lambda\simeq 0.7246$, where the prior distribution of the random parameters are assumed to be uniform with large width windows. 
For a large $d$, the previous expression is always larger than the latter on in the 
braces. Thus, the $\mathcal{O}(d)$ advantage vanishes if the parameter underestimation is a random variable. 
However, this bound is still better than that given by independent estimations~\cite{zhang}. 
Here, we obtain the quantum Ziv-Zakai bound for generalized ECS. 
Through some straightforward calculations, the bound for linear parametrization process is 
\begin{equation}
|\delta \bm{\hat{\theta}}|^2\geq \mathrm{max}\left\{\frac{d(d+\sqrt{d})^2 }{80\lambda^2 (|\alpha|^2+1)^2}, 
\frac{(\pi^2/16-0.5)d(d+\sqrt{d})^2}{(d+\sqrt{d}-1) (|\alpha|^2+1)^2}\right\}.
\end{equation}
Similarly with the generalized NOON state, the $\mathcal{O}(d)$ advantage vanishes in this bound.
However, for a not very small value of $|\alpha|$, $N_{\mathrm{tot}}\simeq|\alpha|^2$, this bound is still lower
than Eq.~(\ref{eq:ZZ_NOON}), which means even for random variables, the generalized ECS can provide a better 
precision than generalized NOON state.

\section{Conclusion\label{sec:Conclusion}}

In this paper,  we have proposed a generalized form of entangled coherent states and apply them as the
input state of a multi-arm interferometer for estimating multiple phase shifts. 
We have obtained the QCRB on the estimation error for both linear and nonlinear protocols. 
Similarly with the generalized NOON state, the simultaneous  estimation with generalized entangled coherent state can provide 
a better precision than the independent estimation. Meanwhile,  We find that  the bound from the 
generalized entangled coherent state is better than that given by the generalized NOON state. 


\ack
The authors thank Heng-Na Xiong and Animesh Datta for valuable discussion. 
This work was supported by the NFRPC 
through Grant No. 2012CB921602 and the NSFC through Grants No. 11475146.
X.-M. L. also acknowledges the support from the Singapore National Research Foundation 
under NRF Grant No. NRF-NRFF2011-07 and NSFC under Grant No. 11304196. 
Z. S. also acknowledges the support from NSFC with grants No. 11375003, 
Natural Science Foundation of Zhejiang Province with grant No. LZ13A040002 and 
Program for HNUEYT with grants No. 2011-01-011.

\end{document}